# RetinaVR: Democratizing Vitreoretinal Surgery Training with a Portable and Affordable Virtual Reality Simulator in the Metaverse

**Fares Antaki**, MDCM, FRCSC[1,2,3], **Cédryk Doucet**, MASc[4], **Daniel Milad**, MD[2,3], **Charles-Édouard Giguère**, MSc[5], **Benoit Ozell**, PhD[4], **Karim Hammamji**, MD, FRCSC[2,3,*]

1. The CHUM School of Artificial Intelligence in Healthcare, Montreal, Quebec, Canada; 2. Department of Ophthalmology, Université de Montréal, Montreal, Quebec, Canada; 3. Department of Ophthalmology, Centre Hospitalier de l'Université de Montréal, Montreal, Quebec, Canada; 4. Department of Computer Engineering and Software Engineering, Polytechnique Montréal, Chemin de Polytechnique, Montréal, Canada; 5. Institut universitaire en santé mentale de Montréal (IUSMM), Montreal, Quebec, Canada. *Correspondence. karim.hammamji@gmail.com

**Funding:** This project received the Innovation in Retina Research Award, including the First Prize (CAD$35,000) and the Audience Award (CAD$5,000), at the COS Annual Meeting in June 2021, co-sponsored by the COS and Bayer Inc.

**We developed and validated RetinaVR, an affordable and immersive virtual reality simulator for vitreoretinal surgery training, using the Meta Quest 2 VR headset. We focused on four core fundamental skills: core vitrectomy, peripheral shaving, membrane peeling, and endolaser application. The validation study involved 10 novice ophthalmology residents and 10 expert vitreoretinal surgeons. We demonstrated construct validity, as shown by the varying user performance in a way that correlates with experimental runs, age, sex, and expertise. RetinaVR shows promise as a portable and affordable simulator, with potential to democratize surgical simulation access, especially in developing countries.**

# Introduction

Virtual reality (VR) simulation in healthcare has made significant progress over the past two decades and is now considered a cornerstone of medical education.[1] In surgery, it enables trainees to acquire skills in an immersive learning environment that mitigates patient harm. The digital nature of VR also alleviates the ethical and logistic challenges tied to wet lab training, while offering an interactive, high-fidelity experience.[2] In ophthalmology, VR simulation has been shown to improve the performance of novice cataract surgeons and to decrease their complication rate.[3,4] Similar trends have been observed in vitreoretinal surgery training, but without definite evidence on skill transfer to the operating room.[5,6]

The most frequently studied VR simulator in ophthalmology is the EyeSi Surgical Simulator (Haag-Streit Simulation). It comprises a mannequin head, surgical instruments, foot pedals, and a VR interface, accessible through the operating microscope.[7] Despite its high cost of acquisition (approximately USD$200,000) and its annual running costs, the use of EyeSi has been shown to be cost-effective for cataract surgery training when considering the reduction of complications.[4,8,9] However, in developing nations and under-resourced communities, the simulator's cost could pose a significant acquisition barrier. This may

disproportionately affect these already vulnerable groups, further exacerbating their risk of adverse health outcomes.[10]

Since the 1970s, head-mounted displays (VR headsets) have steadily decreased in weight and improved in computing capacity. VR headsets have moved beyond academic labs and are commercially available with prices starting from USD$299.[11] VR headsets offer several benefits over traditional stationary simulators, including portability, improved immersiveness, and multiplayer capabilities through 'the metaverse'.[12] This allows multiple users to concurrently use the system and interact together in a virtual environment. By leveraging their existing hardware and software capabilities, VR headsets can democratise access to surgical simulation, making the metaverse a particularly useful space for global ophthalmic education and collaboration.

In this work, we developed a VR simulation application software for vitreoretinal surgery training that is compatible with commercially available VR headsets. RetinaVR is fully immersive, affordable, and portable, as it leverages the powerful processors, cameras, and sensors of the headset without the need for external haptic devices. We focus on four fundamental skills: core vitrectomy, peripheral shaving, membrane peeling, and endolaser application. To our knowledge, this is the first vitreoretinal surgery simulator of its kind

# Methods

We provide an overview of RetinaVR in **Figure 1**. RetinaVR was developed as a simulation app that is compatible with off-the-shelf hardware. We focused our work on the affordable Meta Quest 2 VR headset (Meta Platforms Inc., California), the best-selling VR headset available at the time.[13] Four training modules were built to simulate fundamental skills in vitrectomy surgery.

## Virtual reality hardware

We carried out all development experiments on the wired HP Reverb, attached to an AMD Ryzen 5 computer with 2600x CPU, 16GB of RAM, and an AMD Radeon RX 5700 XT graphics card. After each version iteration, we adapted the app for the wireless Meta Quest 2 to allow our domain experts to test the software remotely and to provide iterative feedback. To ensure broad applicability, we utilized the standard controllers packaged with the Meta Quest 2 only, rather than exploring add-on external haptic devices.

The Meta Quest 2 is a general-purpose VR headset that allows for a standalone experience, eliminating the need for wiring or a computer connection. This feature renders it apt for surgical simulation training, providing an unencumbered environment conducive to learning. It comes with two light-weight plastic controllers, each weighing approximately 150 grams, that are tracked by the headset's integrated cameras. The controllers are designed to rest within the curve of the user's palm, allowing the user's fingers to engage with the capacitive face, grip and trigger buttons as well as the joystick.

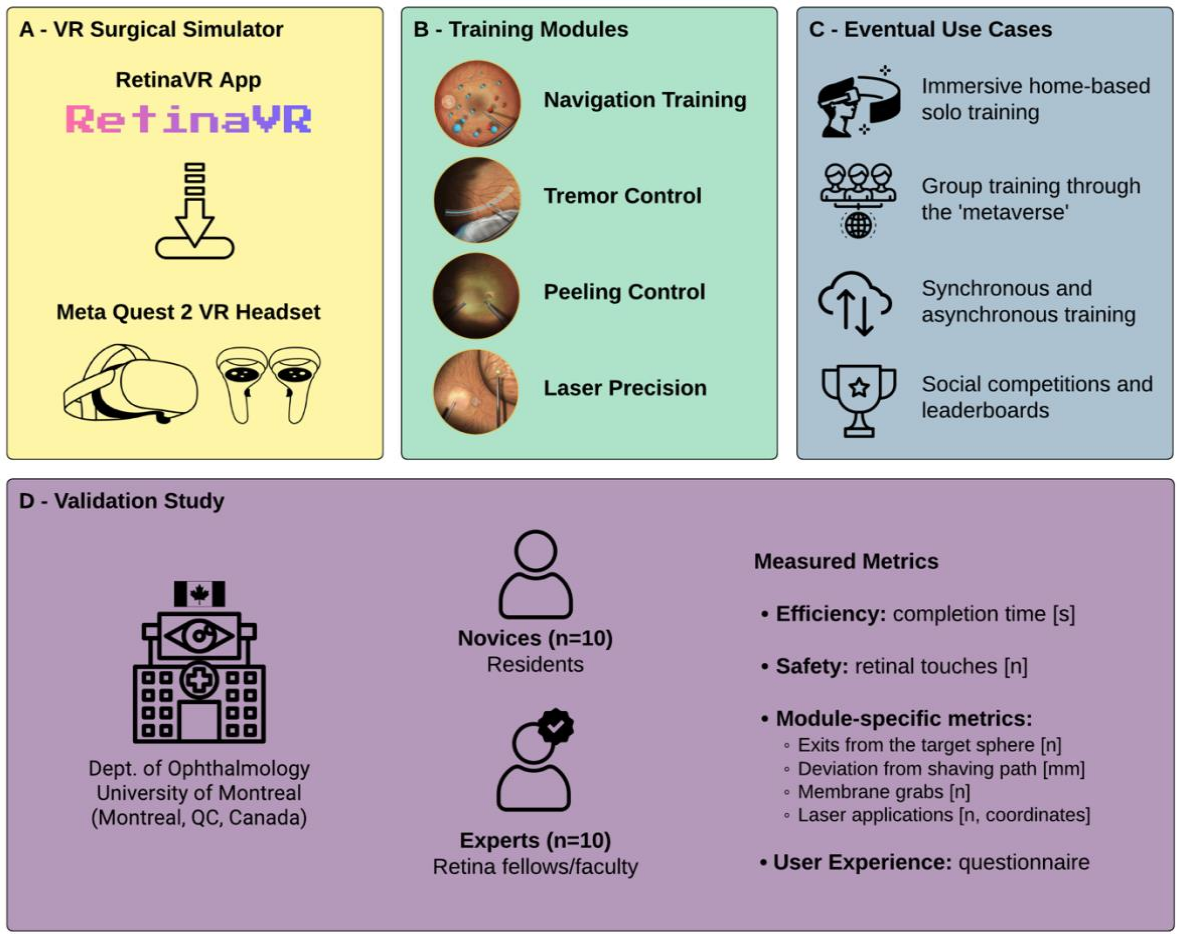

**Figure 1. Overview of the RetinaVR development and validation framework. A.** RetinaVR was developed in the Unity 3D game engine and deployed as an 'app' on the Meta Quest 2 VR headset. **B.** Four training modules simulating fundamental skills in vitrectomy surgery were developed: core vitrectomy (Navigation Training), peripheral shaving (Tremor Control), membrane peeling (Peeling Control), and endolaser application (Laser Precision). **C.** Multiple potential use cases were considered as rationale for selecting the app format and the standalone VR headset. Those included the possibility for home-based solo training, synchronous and asynchronous group training through the metaverse, and social competitions and score leaderboards. **D.** To determine construct validity, we designed a prospective validation study comparing the performance of novice (n=10) and expert users (n=10) recruited from the University of Montreal in Montreal, Quebec, Canada. We analyzed numerous metrics including efficiency, safety and module-specific performance, in relation to their level of expertise and demographic factors.

# Virtual reality software

We developed RetinaVR in the Unity 3D game engine. To represent the eye, a virtual sphere was created, and a custom-made fundus illustration was fitted on its inner surface. The virtual instruments (light pipe and vitrector/ endolaser) were controlled using standard controllers without the use of a physical eye model. The fulcrum effect was challenging to reproduce due to the lack of haptic feedback from the virtual eye and the disconnect between the two controllers. As such, only one controller could be used to move the eye. For all tasks, the left controller was used as a light pipe, while the right controller served as a vitrector or an endolaser probe, and controlled eye movements. The virtual instruments position and their movements were rotated 45 degrees on the x-axis to allow for ergonomic holding of the controllers. To enhance the realism of the simulation, we added the characteristic sound emission produced by the pneumatic guillotine cutter (recorded at 7,500 cuts per minute) during core vitrectomy and peripheral shaving.[14] We also added a laser sound to the endolaser application module. The simulation ergonomics are shown in **Supplemental Figure 2.**

## Training modules

We focused on four fundamental vitreoretinal surgery skills to devise four corresponding training tasks: core vitrectomy (Navigation Training), peripheral shaving (Tremor Control), membrane peeling (Peeling Control), and endolaser application (Laser Precision). Screenshots from each of the modules are shown in **Figure 3**.

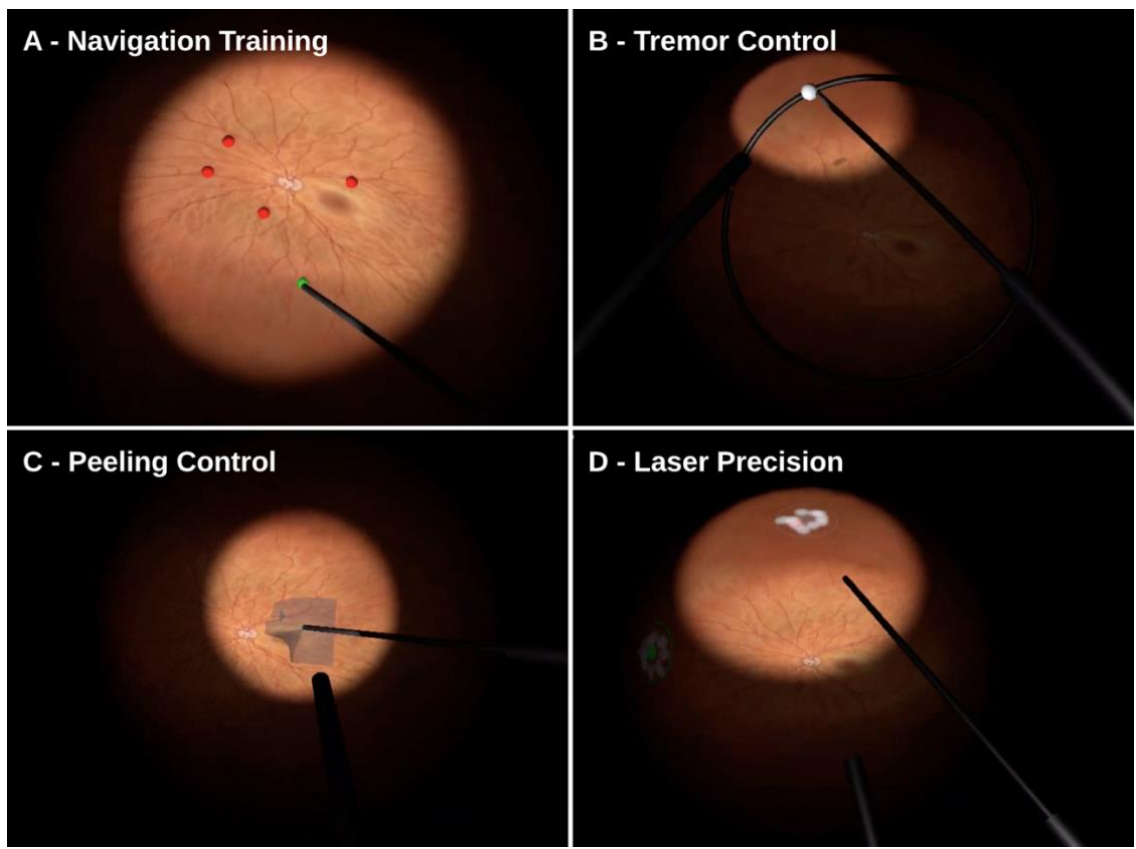

**Figure 3. In-game screenshots from the RetinaVR modules. A.** Navigation Training simulates core vitrectomy. The goal of the user is to collide with all red spheres, maintain the vitrector in the sphere and turn them green. **B.** Tremor Control simulates peripheral shaving. The user will engage the tip of the vitrector with a sphere, allowing it to move along a pre-determined path. **C.** Peeling Control simulates membrane peeling. The user will grasp the membrane by pressing the grip button on the controller before peeling it away from the macula. **D.** Laser Precision simulates endolaser application. The user is asked to treat five retinal breaks by applying laser spots to a surrounding donut. A green marker will indicate a fully-treated tear.

1. **Navigation Training:** To assess navigation skills in the vitreous body, a sphere collection exercise was designed using the 'Collision detection' module in Unity. Initially red, the spheres turn green when collected. To collect a sphere, the tip of the vitrector must maintain contact with it for 2 consecutive seconds (determined heuristically) for it to disappear. The exercise concludes once all 10 spheres at varying depths within the vitreous body are collected.

2. **Tremor Control:** The user's ability to control the vitrector during peripheral shaving is assessed by moving a target sphere along a pre-determined path. When the tip of the vitrector collides with the sphere, it causes the sphere to move along the path until the instrument loses contact with the sphere. The goal is to move the sphere along the path, without deviating, as smoothly as possible without touching the retina.

3. **Peeling Control:** This exercise simulated peeling epiretinal membranes using a cutter-based approach (rather than forceps).[15] The objective was to peel the membrane completely from the retina without iatrogenic touch. Users could enlarge their view by pressing the 'X' button on the left controller, simulating a magnifying lens. To grab the membrane, users were required to press the right grip button. The membrane could only be peeled if a neighboring border was detached, requiring multiple grasps.

4. **Laser Precision:** This exercise focused on applying endolaser around 5 retinal breaks in the periphery. The laser probe had a traditional red spot that varied in size based on its distance to the retina. As in real life, this affected the laser uptake, with larger spots being less intense. The laser was applied by pressing the grip button. Repeat mode was available by holding the button, with an interval of 200ms. When a tear was considered fully-treated, it turned green, signalling to the user to move on to the next break. During development, to ensure that tears were fully treated, we used a raycasting approach in Unity and heuristically adjusted the threshold for "fully treated" until we achieved our desired goal of two rows of laser spots 360 degrees around each break.

## Validation study

After 2 years of development, we locked RetinaVR in March 2023 to prepare it for human validation. Novices and experts were recruited from the Department of Ophthalmology of the University of Montreal in Quebec, Canada from April 2023 through October 2023. The 'Novice' group included ophthalmology residents in their first, second or third years of residency and who have not had any hands-on exposure to intraocular surgery. Exposure to oculoplastic and strabismus surgery, and previous VR exposure (other than RetinaVR) were not exclusionary. The 'Expert' group included experienced fellowship-trained vitreoretinal surgeons and vitreoretinal surgery fellows. The sample size could not be determined *a priori* and was limited by the availability of vitreoretinal surgeons. We recruited all vitreoretinal surgeons at our institution and matched them with an equal number of novice participants. Our sample size is on par with most studies looking at VR simulation in vitreoretinal surgery.[5] We excluded participants if they had any contraindications for VR gaming, including seizure disorder, vertigo, motion sickness and known VR cybersickness. We obtained ethics approval by the Institutional Review Boards of the CHUM Hospital (IRB # 2023-10479-22.035). Informed consent was obtained from all participants after detailing the nature of the study.

Novices and experts were scheduled to test the portable RetinaVR simulator on a Meta Quest 2 headset at their convenience. We often tested in a conference room, requiring only a flat surface. Our lead technical and clinical experts were available during testing, casting the user's view to a connected computer. Before each recorded test, users received a brief explanation of the tasks while wearing the VR headset. They were also instructed on how to position their hands and calibrate the instruments, and they were allowed a single trial run of each module. A life-size soft silicone doll head simulated the patient head, allowing users to rest their wrists. All users sat superiorly relative to the eye.

## Collected data

To determine the construct validity of our simulator, we needed to study the impact of user factors like age, self-reported sex and level of expertise on simulation performance. We collected all possible measurable performance metrics directly from RetinaVR, using built-in code. All modules were evaluated based on three criteria: Efficiency, Safety, and Module-specific performance. For all modules, Efficiency was assessed by measuring completion time in seconds, while Safety was assessed by counting the number of iatrogenic retinal touches. Module-specific performance metrics varied depending on the module. In Navigation Training, the number of exits from the target sphere was counted. For Tremor Control, the number of exits from the target sphere was counted, along with the mean and maximum deviation from the shaving path in milimeters. In Membrane Peeling, the number of membrane grasps was counted, with the hypothesis that the number of grasps would vary with experience and technique. For Laser Precision, the number of laser spots was recorded, with the hypothesis that a parsimonious use of laser was better as long as the tears were treated.[16] The precise coordinates of the laser spots around tears were also recorded to determine the treatment pattern.

## User experience

To measure user experience (UX), we administered a French abbreviated version of the validated Immersive Virtual Environments Questionnaire (IVEQ) v2.[17] The questionnaire consisted of 26 questions: 2 to gauge the user's prior experience with VR, 21 that were gradable using a 10-point Likert scale, and 3 open-ended questions for general comments and feedback. The gradable questions assessed a broad range of UX factors, including *Presence* (n=3), *Engagement* (n=2), *Immersion* (n=2), *Flow* (n=2), *Emotion* (n=2), *Skill* (n=2), *Judgement* (n=3), *Experience Consequence* (n=2), and *Technology Adoption* (n=3). The questionnaire is available in **Appendix 1.** The 3 open-ended questions aimed to gather positive feedback, negative feedback, and suggestions for improvement. To analyze the free text responses, we elucidated the prevalent themes from each response and then consolidated them into broad categories. Once a coherent representation of data across all participants was achieved, the frequency of each theme recorded and summarised.

## Statistical methods

We first explored the unadjusted differences in performance between novices and experts by calculating the standardized mean difference for each performance metric. This effect size analysis was useful to contextualize our findings, given the disparate units (count-, time-, distance-based) and varying scales of the performance metrics stemming from the differing difficulty levels of the four training modules. Additionally, given the novelty of our experimental design, the lack of existing normative data to define "good or bad" or "fast or slow" performance also necessitated this scaled analysis. We interpreted the effects as follows: 0.01 – 0.19 (minimal), 0.20 – 0.49 (small/ mild), 0.50 – 0.79 (medium/ moderate), 0.80 – 0.99 (large) and >1.0 (very large).[18]

We then carried out an adjusted analysis and explored differences between novices and experts while controlling for age, sex and experimental run – factors that can influence the VR gaming experience.[19–21] We used a linear mixed-effect model, which allowed us to isolate the effect of each factor while controlling for all others. All analyses were performed in R V.4.3.1 for our analyses at a 5% alpha level.

# Results

## Baseline demographics

We recruited 20 participants, including 10 novices and 10 experts. Their baseline and demographic characteristics are detailed in **Table 1.** Novices were significantly younger and predominantly female. Novices had no prior surgical experience (a selection criterion), whereas the experts, on average, had 16.6 years (10.71) of post-residency surgical experience. Novices reported more hours of VR gaming than experts, but this difference was not statistically significant. They also reported more hours of training on VR-based surgical simulators than experts, with this difference being statistically significant. We provide descriptive statistics on the performance of novices and experts across all runs and modules in **Supplemental Table 2**, **Supplemental Table 3**, **Supplemental Table 4**, and **Supplemental Table 5**.

| Characteristic | Novices (n=10) | Experts (n=10) | Difference |
|---|---|---|---|
| **Age – yr** | 28.2 (3.61) | 47.1 (12.3) | **p=0.001** |
| **Female sex – n** | 7 (70%) | 3 (30%) | p=0.0736 |
| **Surgical expertise – yr** | 0 (0) | 16.6 (10.71) | **p=0.001** |
| **Previous VR gaming – hrs** | 6.7 (12.82) | 2.1 (3.6) | p=0.299 |
| **Previous VR surgical training – hrs** | 22.6 (23.29) | 3.8 (6.53) | **p=0.033** |

**Table 1. Demographic and baseline characteristics of the novice and expert users.** Values are mean (SD) unless otherwise specified. VR, virtual reality

## Impact of the expertise

We first compared the performance of novices and experts using an unadjusted model. The results are summarized in **Figure 4.** The detailed effect size analyses are shown in **Supplemental Table 6**. The linear mixed-effects model results are summarized in **Table 7**.

Regarding efficiency, we found trends that novices were slower than experts, except in membrane peeling. None of those effects were statistically significant when all experimental runs were combined. When examining the runs individually, we found that novices were faster than experts in the first membrane peeling run (very large effect, -1.10 [95%CI: -2.03, -0.14]). In the linear mixed-effects model, when controlling for age, sex and experimental run, the trends were maintained, but we found no statistically significant difference in efficiency between novices and experts in any of the modules.

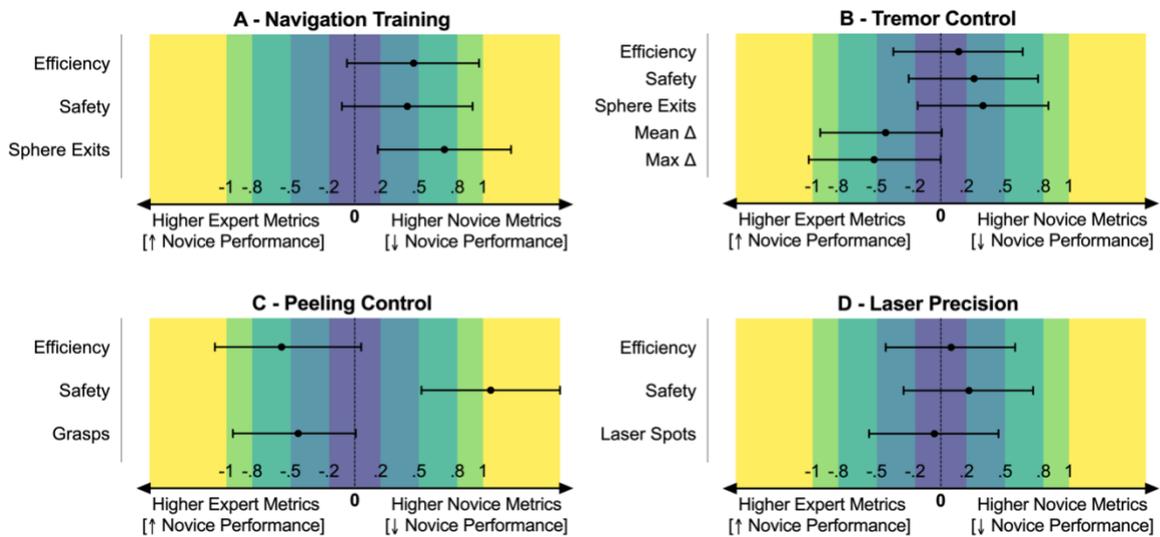

**Figure 4. Forest plot showing the unadjusted effect sizes of efficiency, safety and task-specific performance between experts and novices.** The point effect estimate is Cohen's D and represents the standardized mean difference between novice and expert performance, along with 95% confidence intervals. Positive effects, represented by values to the right of the y-axis, indicate higher novice metrics (e.g., longer novice completion times, more novice retinal touches), suggesting lower novice performance. Conversely, negative effects, represented by values to the left of the y-axis, indicate higher expert metrics (e.g., longer expert completion times, more expert retinal touches), implying better novice performance. The viridis color palette is used to interpret the effect sizes, representing a minimal effect (0.01-0.19), a small or mild effect (0.20-0.49), a medium or moderate effect (0.50-0.79), a large effect (0.80-0.99), and a very large effect (>1.0).

| Module and metric | Expertise | |
|---|---|---|
| | **Estimate** | **p-value** |
| **Navigation Training** | | |
| Efficiency | 9.30 | 0.50 |
| Safety | 0.63 | 0.50 |
| Sphere Exits | 21.46 | **0.014** |
| **Tremor Control** | | |
| Efficiency | 7.05 | 0.29 |
| Safety | 0.10 | 0.94 |
| Sphere Exits | 38.14 | 0.29 |
| Mean Δ | 0.00 | 0.81 |
| Max Δ | -0.02 | 0.71 |
| **Membrane Peeling** | | |
| Efficiency | -16.42 | 0.17 |
| Safety | 2.90 | 0.11 |
| Grasps | -1.03 | 0.25 |
| **Laser Precision** | | |
| Efficiency | 16.68 | 0.29 |
| Safety | 0.63 | 0.29 |
| Laser Spots | 10.15 | 0.64 |

**Table 7. Linear mixed-effects model (adjusted) for the impact of expertise on performance.** This model controls for experimental run, user age and sex. The only significant effect is the difference in performance during Navigation Training. Novices had an excess of 21.46 sphere exists compared to experts (p = 0.014). Efficiency estimates are in seconds, and Safety estimates are in number of iatrogenic retinal touches. Sphere exits, number of grasps and laser spots are count data. Mean and maximal deviation metrics are provided in meters in this table.

Regarding safety, we found that experts were safer than novices in the membrane peeling module when all experimental runs were combined (very large effect, 1.06 [95%CI: 0.52, 1.60]). This effect was also present in the first (very large effect, 1.34 [95%CI: 0.35, 2.31]) and second run (very large effect, 1.12 [95%CI: 0.16, 2.05], but not the third run. We also found trends that the experts were safer in all other modules, but those differences were not statistically significant. In the linear mixed-effects model, when controlling for age, sex and experimental run, the trends were maintained, but we found no statistically significant difference in safety between novices and experts in any of the modules.

Regarding task-specific performance, we found that experts performed better in the core vitrectomy module, demonstrating significantly fewer exits from the target spheres (moderate effect, 0.7 [95%CI: 0.18, 1.22]). This effect was mostly driven by the second experimental run (very large effect, 1.11 [95%CI: 0.15, 2.04]). In the linear mixed-effects model, that difference was maintained while controlling for all other user factors, with novices exiting the spheres an excess of 21.46 times (p = 0.014). In peripheral shaving, we found trends that novices had more sphere exits than experts while demonstrating less deviation from the shaving path, but those differences were not statistically significant. In the linear mixed-effects model, when controlling for other factors, those differences were also not statistically significant.

In membrane peeling, we found trends that experts grasped the membrane more times than novices, but that difference was not statistically significant in the unadjusted model. The trend was maintained in the linear mixed-effects model, but the difference was not statistically significant. In endolaser application, we found no difference in the amount of laser used between novices and experts in the adjusted and unadjusted models. However, a heatmap analysis of the laser spot distribution showed clinically significant differences in treatment patterns among novices and experts, as shown in **Figure 5**.

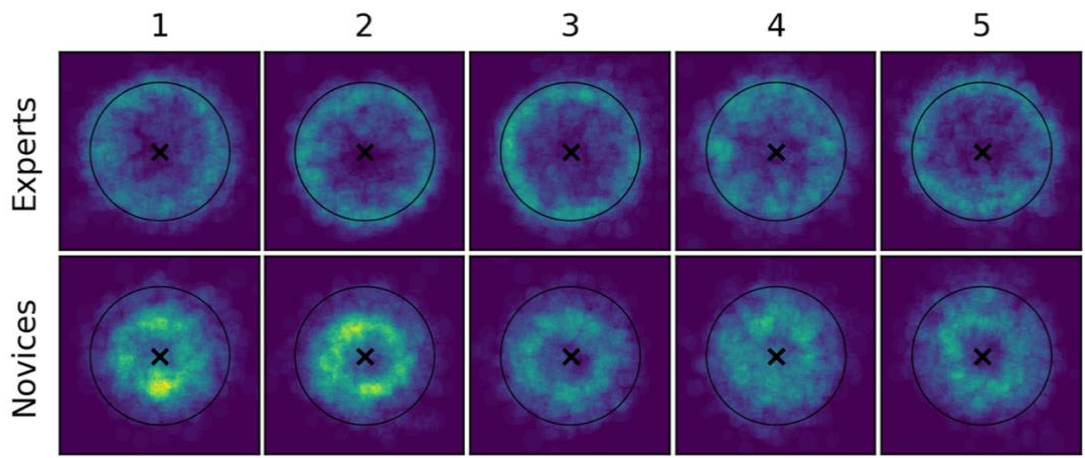

**Figure 5. Heatmap of laser shot distribution in the endolaser application (Laser Precision) module.**
Heatmap illustrating laser treatment patterns for all five retinal tears, differentiated by novices and experts. Each square represents a unique tear, with color intensity corresponding to the number of laser spots applied, using the viridis color palette. The color gradient ranges from purple (least density) to bright yellow (highest density). The central target represents the center-point of the retinal break, as shown in the Laser Precision module. Experts showed a uniform distribution of laser spots, characterized by a consistent spread around each tear, maintaining a uniform distance from the central point. There is a ring-like pattern with minimal laser applications directly on the tears. In contrast, novices showed a more erratic pattern (particularly in tears 1 and 2), with a concentration of laser spots towards the center-point of each tear. This indicates a less controlled application, resulting in a scattered distribution with variable intensity and less discernible uniformity.

## Impact of participant age and sex

We evaluated the impact of participant age and sex on their performance, while controlling for experimental run and expertise. As shown in **Table 8**, in the linear mixed-effects model, age had no impact on performance in any of the modules. Male participants were 12.35 seconds faster in peripheral shaving (p=0.036) and 32.21 seconds faster in membrane peeling (p=0.004) compared to women. We observed trends of males being more efficient, safer and performing better in most task-specific metrics, but none of those effects were statistically significant.

| Module and metric | Age | | Sex | |
|---|---|---|---|---|
| | **Estimate** | **p-value** | **Estimate** | **p-value** |
| **Navigation Training** | | | | |
| Efficiency | 0.29 | 0.64 | -18.96 | 0.11 |
| Safety | 0.00 | 0.93 | -0.17 | 0.83 |
| Sphere Exits | 0.72 | 0.06 | -12.71 | 0.07 |
| **Tremor Control** | | | | |
| Efficiency | 0.52 | 0.09 | -12.35 | **0.036** |
| Safety | 0.00 | 0.94 | -1.13 | 0.32 |
| Sphere Exits | 2.11 | 0.20 | -57.02 | 0.07 |
| Mean Δ | 0.00 | 0.63 | 0.00 | 0.95 |
| Max Δ | 0.00 | 0.54 | 0.00 | 0.94 |
| **Membrane Peeling** | | | | |
| Efficiency | 0.70 | 0.19 | -32.21 | **0.004** |
| Safety | -0.03 | 0.68 | -0.54 | 0.71 |
| Grasps | 0.01 | 0.79 | -1.13 | 0.14 |
| **Laser Precision** | | | | |
| Efficiency | 0.99 | 0.17 | -13.28 | 0.32 |
| Safety | 0.00 | 0.91 | 0.61 | 0.23 |
| Laser Spots | 0.78 | 0.43 | -5.90 | 0.75 |

**Table 8. Linear mixed-effects model (adjusted) for the impact of age and sex on performance.** This model controls for all other variables including expertise when examining the effect of age and sex. Age had no impact on performance, but sex did. Male participants were 12.35 seconds faster in peripheral shaving (p=0.036) and 32.21 seconds faster in membrane peeling (p=0.004). Efficiency estimates are in seconds, and Safety estimates are in number of iatrogenic retinal touches. Sphere exits, number of grasps and laser spots are count data. Mean and maximal deviation metrics are provided in meters in this table.

## Impact of the learning curve

We also evaluated the learning curve by repeating the experiments three times for each participant. As shown in **Table 9**, in the linear mixed-effects model, efficiency improved with each experimental run during all modules. At each run, completion time decreased by 7.67 seconds for core vitrectomy (p=0.005), 12.02 seconds for peripheral shaving (p<0.001), 17.92 seconds for membrane peeling (p<0.001) and 25.68 seconds for endolaser application (p<0.001). We found that repetition improved safety scores during membrane peeling, with a 1.37 fewer iatrogenic retinal touches with each run (p=0.003). Similar trends were observed for all modules, but the effects were not statistically significant. However, it did reduce the number of laser spots used by the participants. At each run, the number of sphere exits decreased by 5.42 times (p=0.038) in core vitrectomy and by 17.00 times during peripheral shaving (p=0.011). In endolaser application, participants used 11.20 less laser shots at each run to treat the tears (p=0.043).

| Module and metric | Experimental run | |
|---|---|---|
| | **Estimate** | **p-value** |
| **Navigation Training** | | |
| Efficiency | -7.67 | **0.005** |
| Safety | -0.20 | 0.26 |
| Sphere Exits | -5.42 | **0.038** |
| **Tremor Control** | | |
| Efficiency | -12.02 | **<0.001** |
| Safety | -0.47 | 0.13 |
| Sphere Exits | -17.00 | **0.011** |
| Mean Δ | 0.00 | 1.00 |
| Max Δ | -0.01 | 0.22 |
| **Membrane Peeling** | | |
| Efficiency | -17.92 | **<0.001** |
| Safety | -1.37 | **0.003** |
| Grasps | 0.27 | 0.23 |
| **Laser Precision** | | |
| Efficiency | -25.68 | **<0.001** |
| Safety | -0.02 | 0.92 |
| Laser Spots | -11.20 | **0.043** |

**Table 9. Linear mixed-effects model (adjusted) for the impact of experimental run on performance.** This model controls for age, sex and expertise when examining the role of the experimental run. At each run, completion time decreased by 7.67 seconds for core vitrectomy (p=0.005), 12.02 seconds for peripheral shaving (p<0.001), 17.92 seconds for membrane peeling (p<0.001) and 25.68 seconds for endolaser application (p<0.001). Furthermore, repetition improved safety scores during membrane peeling, with a 1.37 fewer iatrogenic retinal touches with each run (p=0.003). At each run, the number of sphere exits decreased by 5.42 times (p=0.038) in core vitrectomy and by 17.00 times during peripheral shaving (p=0.011). In endolaser application, participants used 11.20 less laser shots at each run to treat the tears (p=0.043). Efficiency estimates are in seconds, and Safety estimates are in number of iatrogenic retinal touches. Sphere exits, number of grasps and laser spots are count data. Mean and maximal deviation metrics are provided in meters in this table.

## User experience

Overall, the users rated the experience from favorable to excellent in all 8 spheres of UX, as shown in **Supplemental Table 10**. Positive feedback predominantly centered on three themes: the realistic 3D environment (n = 18), the ability to practice in a low-risk environment (n = 9), and the authentic representation of the vitrectomy experience (n = 5). Other sporadic comments praised the innovation, immersion, and portability of the experience. Negative feedback mentioned the fulcrum effect and controller-simulation movement translation (n=8), the controller size and ergonomics (n=6) and difficulty with visualization and depth perception (n=6). Other comments included the lack of progress indicators, headset fit, and unrealistic shaving module. Suggestions for improvement suggested improving the controllers and ergonomics (n=10), providing better instructions and real-time feedback (n=5), and improving movement translation (n=5). It was also recommended to attempt to improve headset fit, build more complete case-based modules, improve graphics, and gamify the experience.

# Discussion

We built a RetinaVR, a fully immersive, affordable, and portable VR simulator for vitreoretinal surgery training. RetinaVR is a standalone app that leverages the powerful processors and cameras of commercially-available VR headsets and controllers, without relying on external touch haptic devices. RetinaVR is a proof of concept for a new way of approaching surgical simulation in the metaverse, at a fraction of the cost of traditional VR simulators. It democratizes access to surgical simulation, and has the potential to spur innovation in global ophthalmology.

To ensure RetinaVR's affordability and accessibility, we designed it to require only a quick app download. The app is a mere 100 megabytes, taking approximately 20 seconds to download on average global broadband speeds and less than 2 minutes in Sub-Saharan Africa.[22,23] To simulate surgical instruments, we used the standard built-in controllers, rather than integrating custom hardware. Using pen-like haptic feedback devices could have provided a more faithful simulation of instruments, but it would have come at a high cost.[24] Since our simulator did not require instruments to be anchored to a physical eye model, the fulcrum effect was difficult to simulate. This effect, encountered when using the vitrector and light pipe through a trocar, necessitates unique skills to move the instrument tips. Despite that, we feel that we accurately replicated the motion inversion and scaled motion required to move the vitrector tip, allowing the users to successfully complete the modules and improve at each run. This is supported by the demonstration of the learning curve and the high scores for the *Flow* theme in the UX questionnaire. The users did suggest, however, improvements in instrumentation. While our plastic controllers were lightweight, they were still considerably heavier than conventional surgical instruments. Their weight was 4 times that of a typical 23G vitrector. For comparison, the Bi-Blade vitrectomy cutter weighs approximately 37 grams with the tubing (personal communication with Bausch + Lomb).

To capture user performance during simulation, we were faced with two options: either collect as many metrics as possible and analyze them post-hoc, or develop a scoring system by assigning weights to measurable metrics based on our subjective assessment of their importance. The latter approach raised concerns about how to objectively measure task efficiency, safety, and good performance, and how to determine the appropriate point deductions for mistakes. Given the potential for heuristic bias, we chose the first option and developed code in RetinaVR to quantify those metrics. We conducted a rigorous analysis of the data through an effect size analysis. This was crucial for interpreting the significance of observed differences, since these experiments were being conducted for the first time with no normative databases to establish good or poor performance benchmarks. We then built an adjusted model to examine the impact of age, sex and experimental run on performance, and controlled for those factors when comparing novices and experts.

We believe to have demonstrated construct validity.[25,26] This refers to the ability RetinaVR to measure user behaviors and performance in a way that correlates with their inherent factors and level of expertise. We found that participant age had no impact on overall performance when we controlled for sex, expertise and experimental run. However, we found that males performed membrane peeling and peripheral shaving tasks more quickly than females, with no significant differences in safety and task-specific performance.

Some evidence suggests that gaming proficiency may decline with age and show differences between sexes.[21,27–29] However, we believe that this phenomenon is more likely attributable to a disparity in prior gaming experience, rather than innate age or sex-related abilities. These effects may be even less pronounced in a surgical simulation context like ours, where older participants typically have more prior surgical experience. In parallel, we found that repetition boosted efficiency in all modules, and enhanced safety in the membrane peeling module. It also improved task-specific performance during core vitrectomy and peripheral shaving. This demonstrates a learning curve across experimental runs – with users getting better with repetition or practice. We feel that this observation reinforces the notion that user performance was not a random occurrence but rather a reflection of genuine learning. This learning curve has also been demonstrated for the vitreoretinal modules of the EyeSi simulator in numerous studies.[30,31]

A crucial aspect of this project is the demonstration of how user expertise affects performance. We report on several notable findings in our work. First, novices tended to be slower in all modules, except in membrane peeling. Interestingly, in membrane peeling, they tended to be faster, while also being less safe, causing significantly more iatrogenic retinal touches, and grasping the membranes less frequently. These contrasts possibly highlight the influence of real-world surgical experience. Experts demonstrated a more cautious and deliberate approach, peeling slowly and carefully to minimize shearing forces on the macula. In contrast, novices, perhaps viewing the simulation as such, exhibited riskier behavior by attempting to complete the module at a faster pace, leading to more iatrogenic damage. Second, experts performed significantly better in the core vitrectomy module, exhibiting fewer target sphere exits – a difference that was maintained when controlling for other user factors. Third, during endolaser application, we found clinically important differences in the treatment patterns between among novices and experts. This speaks to the construct validity of those modules and their ability to faithfully simulate the surgical experience.

RetinaVR marks a proof of concept for a novel type of platform for vitreoretinal surgery training simulation. We believe that RetinaVR can change the scope of surgical simulation in a number of ways. First, trainees can conveniently access RetinaVR using their personal headsets, integrating it alongside their existing VR-based entertainment, gaming, or sports activities. Second, residency programs can effectively train multiple residents simultaneously by investing in multiple affordable VR headsets. The platform's online metaverse integration, relying on Meta's cloud servers, enables multiplayer group training sessions, connecting residents virtually with expert surgeons from around the world, breaking down geographical barriers and fostering a global learning community. Third, the platform allows for both synchronous and asynchronous learning, which enables trainees to obtain real-time feedback from mentors while also catering for individual learning styles and schedules. Finally, gamification elements, such as points, badges, and international leaderboards, can further enhance engagement and encourage healthy competition, spurring innovation and collaboration in the field of vitreoretinal surgery.

While RetinaVR has demonstrated construct validity to a certain extent, our work has some limitations and further validation is necessary. First, statistical significance in our analyses was limited by the low sample size and high variance among novices. Despite that, most of our effects were congruent with the expected behaviors of novices and experts. Second, we have not yet demonstrated skill transfer to the operating room, a crucial step in validating a surgical simulator. However, to our knowledge, in vitreoretinal surgery, this has not been shown even for popular simulators like the EyeSi.[5] RetinaVR remains a work in progress: the user interface, including menu appearances, profile creation, login functionality, and leaderboards, require further development before public release. We are also working on incorporating feedback from this study to determine future directions for RetinaVR. Despite these limitations, we are proud of what was achieved with limited resources. RetinaVR serves as a proof of concept for developing affordable VR surgical simulation apps in an academic lab setting, fostering innovation in surgical training and medical education. Driven by the relentless innovation of industry titans like Meta and Apple, we are confident that standalone VR headsets will soon reach a high level of maturity.[32] This will pave the way for the widespread availability of an off-the-shelf, affordable, and validated RetinaVR app, empowering the trainees worldwide with an immersive surgical training experience.

# Statements

**Contributions:** FA, CD, BO and KH conceptualised the study and designed the experiments. FA and KH obtained the funding. CD and BO designed RetinaVR software. FA and KH provided continuous iterative feedback to improve RetinaVR. FA, CD, and DM carried out the clinical validation study. CEG performed the statistical analyses. FA and CD drafted the initial manuscript. FA and CD designed the figures and tables. All authors reviewed and discussed the results. All authors edited and revised the manuscript before approving the final version of this manuscript.

**Acknowledgements:** We express our gratitude to the Canadian Ophthalmological Society and Bayer Inc for funding this work.

**Data sharing statement:** All data produced in the present study are available upon reasonable request to the authors. RetinaVR is not currently in the Oculus store.

# Supplemental Materials

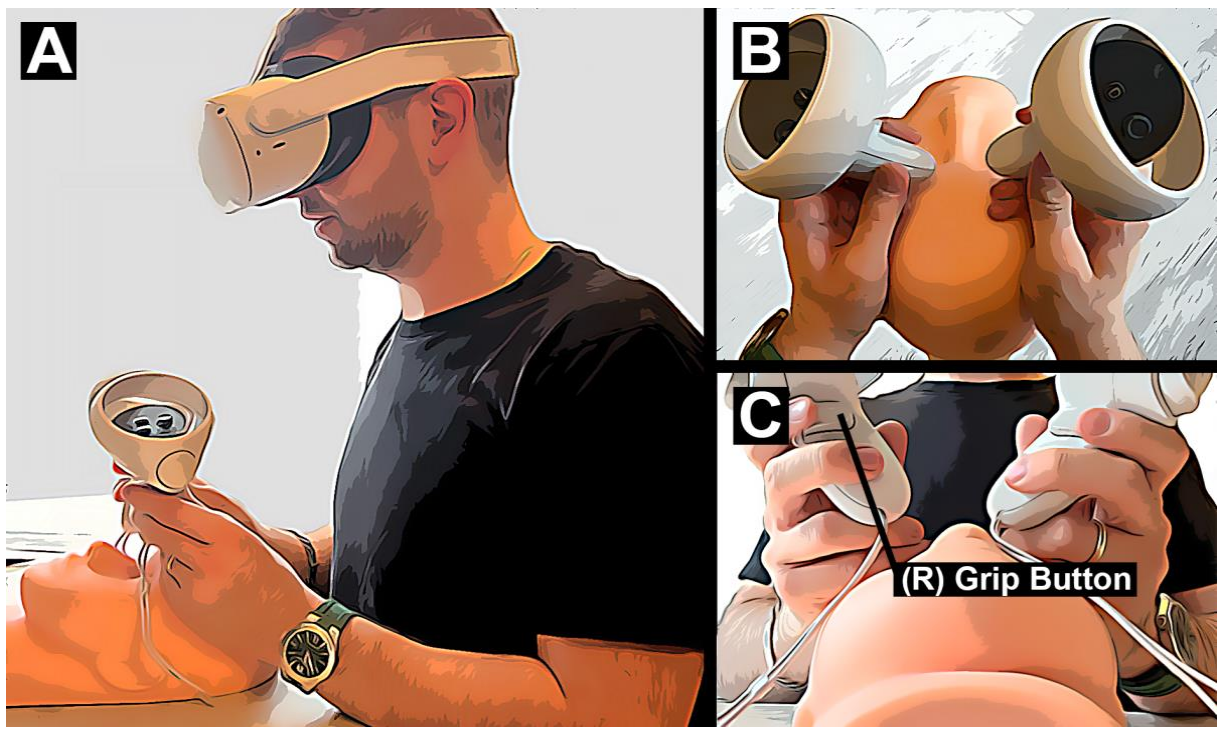

**Supplemental Figure 2. Overview of the simulation ergonomics. A.** At the beginning of each simulation, users could calibrate the position of their instruments to a comfortable position while resting the ulnar border of their hands on a silicon mannequin forehead. **B.** Top-down view from the user's perspective. The controllers are held like surgical instruments. **C.** The right grip button *(Axis1D.PrimaryHandTrigger)* serves two primary functions: firstly, it enables membrane grasping during membrane peeling, and secondly, it engages the laser during endolaser application. Continuous pressing activates repeat laser firing at 200-millisecond intervals.

| Run | 1 | | 2 | | 3 | | Overall | |
|---|---|---|---|---|---|---|---|---|
| | **Expert** | **Novice** | **Expert** | **Novice** | **Expert** | **Novice** | **Expert** | **Novice** |
| **Efficiency [completion time in seconds]** | | | | | | | | |
| **Mean** | 58.59 | 70.41 | 43.59 | 62.57 | 47.4 | 50.9 | 49.86 | 61.3 |
| **StDev** | 17.52 | 41.12 | 6.35 | 34.7 | 12.83 | 15.22 | 14.17 | 32.2 |
| **Min** | 38.45 | 41.48 | 36.98 | 38.37 | 37.56 | 33.65 | 36.98 | 33.65 |
| **Max** | 98.86 | 175.74 | 58.25 | 148.33 | 75.69 | 76.73 | 98.86 | 175.74 |
| **Safety [number of iatrogenic retinal touches]** | | | | | | | | |
| **Mean** | 0.6 | 1.4 | 0.4 | 0.9 | 0.3 | 0.9 | 0.43 | 1.07 |
| **StDev** | 1.58 | 2.46 | 0.7 | 1.85 | 0.67 | 1.52 | 1.04 | 1.93 |
| **Min** | 0 | 0 | 0 | 0 | 0 | 0 | 0 | 0 |
| **Max** | 5 | 8 | 2 | 6 | 2 | 5 | 5 | 8 |
| **Performance [number of exits from the target sphere]** | | | | | | | | |
| **Mean** | 10.6 | 35 | 7 | 16.6 | 9.5 | 14.4 | 9.03 | 22 |
| **StDev** | 5.6 | 39.98 | 5.4 | 11 | 8.15 | 8.32 | 6.46 | 25.36 |
| **Min** | 2 | 8 | 2 | 6 | 3 | 5 | 2 | 5 |
| **Max** | 21 | 109 | 17 | 38 | 31 | 35 | 31 | 109 |

**Supplemental Table 2. Descriptive statistics for Navigation Training**

| Run | 1 | | 2 | | 3 | | Overall | |
|---|---|---|---|---|---|---|---|---|
| | Expert | Novice | Expert | Novice | Expert | Novice | Expert | Novice |
| **Efficiency [completion time in seconds]** | | | | | | | | |
| **Mean** | 55.74 | 58.57 | 38.56 | 42.61 | 33.22 | 33.02 | 42.5 | 44.73 |
| **StDev** | 20.97 | 13.6 | 11.63 | 11.58 | 7.77 | 5.4 | 17.11 | 14.93 |
| **Min** | 29.49 | 36.88 | 25.99 | 24.64 | 23.15 | 22.22 | 23.15 | 22.22 |
| **Max** | 90.65 | 77.27 | 61.8 | 62.62 | 43.94 | 39.19 | 90.65 | 77.27 |
| **Safety [number of iatrogenic retinal touches]** | | | | | | | | |
| **Mean** | 1.1 | 1.6 | 1.1 | 2.3 | 0.3 | 0.5 | 0.83 | 1.47 |
| **StDev** | 1.45 | 2.91 | 1.85 | 4.69 | 0.48 | 0.97 | 1.39 | 3.21 |
| **Min** | 0 | 0 | 0 | 0 | 0 | 0 | 0 | 0 |
| **Max** | 4 | 9 | 6 | 13 | 1 | 3 | 6 | 13 |
| **Performance [number of exits from the target sphere]** | | | | | | | | |
| **Mean** | 168.9 | 222.7 | 183.1 | 186 | 158.6 | 165 | 170.2 | 191.23 |
| **StDev** | 47.25 | 62.43 | 66.33 | 86.27 | 48.13 | 64.3 | 53.68 | 73.42 |
| **Min** | 86 | 147 | 115 | 97 | 110 | 94 | 86 | 94 |
| **Max** | 232 | 336 | 305 | 396 | 230 | 288 | 305 | 396 |
| **Performance [mean Δ from the shaving path in mm]** | | | | | | | | |
| **Mean** | 19.7 | 16.8 | 25.2 | 16.0 | 21.3 | 15.2 | 22.0 | 16.0 |
| **StDev** | 6.6 | 1.2 | 28.4 | 1.1 | 19.8 | 0.6 | 19.8 | 1.1 |
| **Min** | 14.3 | 15.0 | 13.9 | 14.1 | 13.4 | 13.9 | 13.4 | 13.9 |
| **Max** | 34.9 | 18.5 | 105.2 | 17.8 | 77.6 | 15.8 | 105.2 | 18.5 |
| **Performance [maximal Δ from the shaving path in mm]** | | | | | | | | |
| **Mean** | 98.3 | 39.8 | 110.5 | 44.1 | 64.1 | 30.3 | 91.0 | 38.1 |
| **StDev** | 94.4 | 18.0 | 209.9 | 26.8 | 104.8 | 8.6 | 142.3 | 19.5 |
| **Min** | 18.0 | 19.9 | 18.6 | 22.6 | 17.4 | 21.8 | 17.4 | 19.9 |
| **Max** | 305.6 | 76.1 | 692.0 | 106.3 | 357.4 | 52.9 | 692.0 | 106.3 |

**Supplemental Table 3. Descriptive statistics for Tremor Control**

| Run | 1 | | 2 | | 3 | | Overall | |
|---|---|---|---|---|---|---|---|---|
| | **Expert** | **Novice** | **Expert** | **Novice** | **Expert** | **Novice** | **Expert** | **Novice** |
| **Efficiency [completion time in seconds]** | | | | | | | | |
| **Mean** | 97.92 | 65.49 | 56.86 | 50.58 | 51.58 | 40.15 | 68.79 | 52.07 |
| **StDev** | 30.43 | 28.49 | 22.15 | 24.97 | 17.52 | 23.37 | 31.28 | 26.96 |
| **Min** | 43.23 | 43.75 | 30.88 | 25.02 | 24.64 | 21.95 | 24.64 | 21.95 |
| **Max** | 132.33 | 135.04 | 90.51 | 106.08 | 82.8 | 100.66 | 132.33 | 135.04 |
| **Safety [number of iatrogenic retinal touches]** | | | | | | | | |
| **Mean** | 1.8 | 6.3 | 0.2 | 4.9 | 0.3 | 2.3 | 0.77 | 4.5 |
| **StDev** | 2.53 | 4 | 0.63 | 5.92 | 0.95 | 3.13 | 1.72 | 4.66 |
| **Min** | 0 | 0 | 0 | 0 | 0 | 0 | 0 | 0 |
| **Max** | 6 | 11 | 2 | 20 | 3 | 10 | 6 | 20 |
| **Performance [number of membrane grasps]** | | | | | | | | |
| **Mean** | 8.9 | 7.6 | 8.7 | 8.7 | 9.3 | 8.3 | 8.97 | 8.2 |
| **StDev** | 2.33 | 1.51 | 1.89 | 1.49 | 1.77 | 1.49 | 1.96 | 1.52 |
| **Min** | 5 | 6 | 6 | 7 | 7 | 6 | 5 | 6 |
| **Max** | 14 | 11 | 12 | 12 | 13 | 11 | 14 | 12 |

**Supplemental Table 4. Descriptive statistics for Peeling Control**

| Run | 1 | | 2 | | 3 | | Overall | |
|---|---|---|---|---|---|---|---|---|
| | **Expert** | **Novice** | **Expert** | **Novice** | **Expert** | **Novice** | **Expert** | **Novice** |
| **Efficiency [completion time in seconds]** | | | | | | | | |
| Mean | 140.54 | 125.87 | 85.7 | 109.53 | 81.61 | 82.08 | 102.62 | 105.82 |
| StDev | 39.64 | 36.2 | 15.21 | 58.42 | 25.56 | 13.32 | 38.85 | 43.11 |
| Min | 71.8 | 76.73 | 55.57 | 40.6 | 47 | 62.57 | 47 | 40.6 |
| Max | 192.61 | 201.48 | 112.71 | 255.54 | 134.98 | 106.5 | 192.61 | 255.54 |
| **Safety [number of iatrogenic retinal touches]** | | | | | | | | |
| Mean | 0.2 | 1 | 0 | 1.1 | 1 | 0.1 | 0.4 | 0.73 |
| StDev | 0.63 | 1.89 | 0 | 1.85 | 2.54 | 0.32 | 1.52 | 1.55 |
| Min | 0 | 0 | 0 | 0 | 0 | 0 | 0 | 0 |
| Max | 2 | 6 | 0 | 5 | 8 | 1 | 8 | 6 |
| **Performance [number of laser spots]** | | | | | | | | |
| Mean | 173.7 | 162.2 | 149.4 | 165.5 | 151.3 | 139.8 | 158.13 | 155.83 |
| StDev | 52.32 | 21.62 | 39.06 | 47.11 | 52.75 | 27.56 | 48.09 | 34.7 |
| Min | 121 | 129 | 99 | 93 | 98 | 98 | 98 | 93 |
| Max | 282 | 194 | 232 | 243 | 250 | 183 | 282 | 243 |

**Supplemental Table 5. Descriptive statistics for Laser Precision**

| Task/ Run | 1 | 2 | 3 | Combined |
|---|---|---|---|---|
| **Navigation Training** | | | | |
| Efficiency | 0.37 [-0.52, 1.25] | 0.76 [-0.16, 1.66] | 0.25 [-0.63, 1.13] | 0.46 [-0.06, 0.97] |
| Safety | 0.39 [-0.50, 1.27] | 0.36 [-0.53, 1.24] | 0.51 [-0.39, 1.39] | 0.41 [-0.10, 0.92] |
| Sphere Exits | 0.85 [-0.08, 1.76] | **1.11 [0.15, 2.04]** | 0.59 [-0.31, 1.48] | **0.7 [0.18, 1.22]** |
| **Tremor Control** | | | | |
| Efficiency | 0.16 [-0.72, 1.04] | 0.35 [-0.54, 1.23] | -0.03 [-0.91, 0.85] | 0.14 [-0.37, 0.64] |
| Safety | 0.22 [-0.67, 1.09] | 0.34 [-0.55, 1.22] | 0.26 [-0.62, 1.14] | 0.26 [-0.25, 0.76] |
| Sphere Exits | **0.97 [0.03, 1.89]** | 0.04 [-0.84, 0.91] | 0.11 [-0.77, 0.99] | 0.33 [-0.18, 0.84] |
| Mean Δ | -0.61 [-1.50, 0.30] | -0.46 [-1.34, 0.43] | -0.43 [-1.32, 0.46] | -0.43 [-0.94, 0.08] |
| Max Δ | -0.86 [-1.77, 0.07] | -0.44 [-1.33, 0.45] | -0.45 [-1.34, 0.44] | -0.52 [-1.03, 0.00] |
| **Peeling Control** | | | | |
| Efficiency | **-1.10 [-2.03, -0.14]** | -0.27 [-1.14, 0.62] | -0.55 [-1.44, 0.35] | -0.57 [-1.09, 0.05] |
| Safety | **1.34 [0.35, 2.31]** | **1.12 [0.16, 2.05]** | 0.87 [-0.07, 1.77] | **1.06 [0.52, 1.60]** |
| Grasps | -0.66 [-1.56, 0.25] | 0 [-0.88, 0.88] | -0.61 [-1.50, 0.30] | -0.44 [-0.95, 0.08] |
| **Laser Precision** | | | | |
| Efficiency | -0.39 [-1.27, 0.50] | 0.56 [-0.34, 1.45] | 0.02 [-0.85, 0.90] | 0.08 [-0.43, 0.58] |
| Safety | 0.57 [-0.33, 1.46] | 0.84 [-0.09, 1.75] | -0.50 [-1.38, 0.40] | 0.22 [-0.29, 0.72] |
| Laser Spots | -0.29 [-1.16, 0.60] | 0.37 [-0.52, 1.25] | -0.27 [-1.15, 0.61] | -0.05 [-0.56, 0.45] |

**Supplemental Table 6. Effect size analysis for all four RetinaVR modules**

| Theme | Expertise | Mean | StDev | Min | Max |
|---|---|---|---|---|---|
| **Presence** | **Expert** | 6 | 2.01 | 1.67 | 8.33 |
|  | **Novice** | 7 | 1.39 | 5.33 | 9.67 |
| **Engagement** | **Expert** | 7 | 1.44 | 5.00 | 8.50 |
|  | **Novice** | 8 | 1.51 | 5.50 | 10.00 |
| **Immersion** | **Expert** | 7 | 2.08 | 3.50 | 9.00 |
|  | **Novice** | 8 | 1.57 | 6.00 | 10.00 |
| **Flow** | **Expert** | 7 | 1.66 | 4.50 | 9.50 |
|  | **Novice** | 8 | 1.38 | 6.00 | 10.00 |
| **Emotion** | **Expert** | 8 | 1.53 | 5.50 | 10.00 |
|  | **Novice** | 8 | 0.97 | 7.00 | 9.50 |
| **Skill** | **Expert** | 7 | 2.24 | 3.00 | 9.00 |
|  | **Novice** | 8 | 1.42 | 5.50 | 9.00 |
| **Judgement** | **Expert** | 8 | 1.62 | 4.33 | 10.00 |
|  | **Novice** | 9 | 1.05 | 6.67 | 10.00 |
| **Experience Consequence** | **Expert** | 8 | 1.96 | 4.00 | 10.00 |
|  | **Novice** | 8 | 2.49 | 2.50 | 10.00 |
| **Technology Adoption** | **Expert** | 9 | 1.71 | 4.67 | 10.00 |
|  | **Novice** | 9 | 0.99 | 7.00 | 10.00 |

**Supplemental Table 10. Results from the User Experience questionnaire**

| Theme | Question |
|---|---|
| Background | Approximately how many hours of experience do you have with virtual reality environments? |
| Background | Approximately how many hours of experience do you have with medical simulation software? |
| Presence | My interactions with the virtual environment seemed natural. |
| Presence | I could concentrate on the assigned tasks rather than on the devices (gamepad or keyboard). |
| Presence | The devices (gamepad or keyboard) which controlled my movement in the virtual environment seemed natural. |
| Engagement | The sense of moving around inside the virtual environment was compelling. |
| Engagement | The visual aspects of the virtual environment involved me. |
| Immersion | I become so involved in the virtual environment that it is if I was inside the game rather than manipulating a gamepad and watching a screen. |
| Immersion | I felt physically fit in the virtual environment |
| Flow | I felt I could perfectly control my actions. |
| Flow | At each step, I knew what to do. |
| Emotion | I enjoyed being in this virtual environment. |
| Emotion | The interaction devices (Oculus headset, gamepad and/or keyboard) bored me to death. |
| Skill | I felt confident using the gamepad and/or keyboard to move around the virtual environment. |
| Skill | I feel confident learning advanced skills within a specific virtual reality software using the Oculus headset. |
| Judgement | Personally, I would say the virtual environment is impractical/practical. |
| Judgement | Personally, I would say the virtual environment is unruly/manageable. |
| Judgement | Personally, I would say the virtual environment is confusing/clear |
| Experience consequence | I suffered from fatigue during my interaction with the virtual environment. |
| Experience consequence | I suffered from dizziness with eye open during my interaction with the virtual environment |
| Technology adoption | It would be easy for me to become skillful at using the virtual environment. |
| Technology adoption | If I use again the same virtual environment, my interaction with the environment would be clear and understandable for me. |
| Technology adoption | Using the interaction devices (Oculus headset, gamepad and/or keyboard) is a bad idea. |
| Feedback | In your opinion, what were the positive points about your experience? |
| Feedback | In your opinion, what were the negative points about your experience? |
| Feedback | Do you have suggestions to improve this virtual reality environment? |

**Appendix 1. User experience questionnaire abridged from the validated IVEQ v2 user experience questionnaire.** The original questionnaire is in French, but the English translation is shown here. Background and feedback questions were open-ended, allowing for natural language inputs. The remaining questions used a 10-point Likert scale. Scores from negatively worded questions were inverted for analysis.